# The $CO_2$–broadened $H_2O$ continuum in the 100-1500 cm$^{-1}$ region. Measurements, predictions and empirical model

Ha Tran[1,#], Martin Turbet[1], Simon Hanoufa[1], Xavier Landsheere[2],
Pascale Chelin[2], Qiancheng Ma[3], Jean-Michel Hartmann[4]

[1] *Laboratoire de Météorologie Dynamique, IPSL, CNRS, Sorbonne Université, École normale supérieure, PSL Research University, École polytechnique, F-75005 Paris, France.*
[#] *Corresponding author: ha.tran@lmd.jussieu.fr*
[2] *Laboratoire Interuniversitaire des Systèmes Atmosphériques (LISA, CNRS UMR 7583). Université Paris Est Créteil, Université Paris Diderot, Institut Pierre-Simon Laplace, 94010 Créteil Cedex, France.*
[3] *NASA/Goddard Institute for Space Studies and Department of Applied Physics and Applied Mathematics, Columbia University, 2880 Broadway, New York, New York 10025, USA.*
[4] *Laboratoire de Météorologie Dynamique/IPSL, CNRS, École polytechnique, Sorbonne Université, École normale supérieure, PSL Research University, F-91120 Palaiseau, France.*

**Abstract**

Transmission spectra of $H_2O$+$CO_2$ mixtures have been recorded, at 296, 325 and 366 K, for various pressures and mixture compositions using two experimental setups. Their analysis enables to retrieve values of the "continuum" absorption by the $CO_2$-broadened $H_2O$ line wings between 100 and 1500 cm$^{-1}$. The results are in good agreement with those, around 1300 cm$^{-1}$, of the single previous experimental study available. Comparisons are also made with direct predictions based on line-shape correction factors χ calculated, almost thirty years ago, using a quasistatic approach and an input $H_2O$-$CO_2$ intermolecular potential. They show that this model quite nicely predicts, with slightly overestimated values, the continuum over a spectral range where it varies by more than three orders of magnitude. An empirical correction is proposed, based on the experimental data, which should be useful for radiative transfer and climate studies in $CO_2$ rich planetary atmospheres.

## 1. Introduction

As explained in [1] and references therein, properly modeling the absorption spectra of $CO_2$+$H_2O$ mixtures under various temperature and pressure conditions as well as for different gas concentrations is crucial for several applications in planetary sciences. More specifically related to the present study, knowledge of the continuum of water vapor diluted in carbon dioxide is needed to accurately model the greenhouse effect due to $H_2O$ in $CO_2$-dominated atmospheres and the associated thermal infrared emission spectrum. The former application is important to better constrain the radiative budget and thus the evolution of atmospheres such as those of the Solar System rocky planets in their early stages, of the present-day Venus, and of extra-solar analogues. The latter application is required to improve our understanding of the thermal emission spectrum of present-day Venus as well as to prepare future observations of the thermal emission of rocky exo-planets, such as the innermost planets of the TRAPPIST-1 system with the forthcoming James Webb Space Telescope.

We here continue the efforts initiated in [1] in order to provide experimental information and models on the continuum absorption by the far line wings of $CO_2$+$H_2O$ mixtures. While [1] studied the effect of humidity (of collisions between $CO_2$ and $H_2O$ molecules) on the wings of $CO_2$ lines, we focus here on the influence of $H_2O$-$CO_2$ intermolecular interactions on the wings of the water vapor lines, i.e. the so-called $CO_2$-broadened $H_2O$ continuum. Despite its importance for planetary studies, this continuum has received very little attention, contrary to the self- and $N_2$- (or air-) broadened continua which have been extensively investigated (see reviews in Refs. [2-4]) due to their role in the Earth atmosphere. To the best of our knowledge, there are only two works devoted to the continuum of $H_2O$ in $CO_2$. The first is a purely theoretical study in which direct predictions were made using a quasistatic approach [5] but could not be tested due to the lack, at that time, of any experimental results. In the second, much more recent, measurements were made around 1300 cm$^{-1}$ [6] but no comparison with predictions of the former model was made.

We here present the results of measurements of the transmission by $H_2O+CO_2$ mixtures made using two different setups altogether covering the spectral range from 40 to 4800 cm$^{-1}$. The recorded data are analyzed in order to retrieve values of the $CO_2$-broadened $H_2O$ continuum at some wave numbers between 100 and 1500 cm$^{-1}$. The results are compared with previous measurements [6] and direct calculations based on $H_2O$ line-shape correction factors $\chi$ predicted [5] using a quasistatic model. For better agreement between measured and computed continua, empirical corrections are made to the $\chi$ factors. The experimental setups, the measurement and data analysis procedures are presented in Sec. 2. The theoretical model for the calculation of the $CO_2$-broadened $H_2O$ continuum is described in Sec. 3. The results are then presented and discussed in Sec. 4 where empirical line shape corrections leading to good agreement with experimental values are also proposed.

## 2. Experimental details and measured spectra analysis
*2.1 Experiments*

The spectra used in the present study were recorded using two different experimental IR facilities.

The first one, located at LISA, is the same as that used in [1]. It includes a (Bruker IFS 120 HR) high-resolution Fourier-transform spectrometer (FTS) equipped with a globar light source, a KBr beam splitter and a MCT detector. An unapodized spectral resolution of 0.1 cm$^{-1}$ was retained, with an iris aperture diameter of 4 mm. A heatable White-type Pyrex glass absorption cell equipped with 4 mm-thick ZnSe windows was connected to the FTS with a dedicated optical inter-face inside the sample chamber. Its base length is 0.20 m and, for the present experiments, an optical path length of 7.20 m was used. This cell can be heated up to 373 K with spatial variations smaller than 0.5 K, as measured with a type-K thermocouple (±1.5 K uncertainty). In order to avoid water-vapor condensation and to work with significant $H_2O$ pressures, the cell and the entire gas-handling system (including the pressure gauges) were enclosed inside a thermally insulated Plexiglas box. The temperature inside this box, regulated by an air heating system, was about 333 K. The gas pressure was measured using two capacitive pressure transducers with 0-0.013 and 0-1.33 bar full scales, with a stated accuracy of ±0.12%. The spectral range from 450-4800 cm$^{-1}$ was recorded for all measurements. The experiments were carried out as follows: Firstly, the temperatures inside the cell and the box were set to the desired values. When these temperatures were stabilized (after about 1 h for the box and 5 hours for the cell), an empty cell spectrum was collected (after pumping out down to a pressure < 10$^{-2}$ mb for which residual absorption do not affect the spectra analysis described in the next section), providing the 100% transmission level. A spectrum of pure $CO_2$ at a pressure of about 1 bar was then recorded. After evacuation, the cell was filled with water vapor, purified by several distillations, at the desired pressure (about 1 hour for filling and waiting equilibrium) and $CO_2$ was introduced until the total pressure reached about 1 bar. Once the sample was well mixed, a series of spectra was collected, progressively reducing the gas sample total pressure, using an averaging of 400 scans. The temperature and pressure inside the cell were simultaneously recorded every 10 s, showing variations during an experiment lower than 0.2 K and 0.7 mb, respectively. In these experiments, total pressures between about 0.5 and 1 bar were investigated, for $H_2O+CO_2$ mixtures containing from 10 to 14 % of water vapor, at 325 and 366 K.

The second setup, located at the AILES beam line of the SOLEIL synchrotron is the same as that used in [7]. It includes a (Bruker IFS 125 HR) high-resolution FTS equipped with a globar light source, a 6 μm-thick mylar beam splitter, a band-pass filter, and a Si bolometer detector cooled down to 4.2 K by liquid helium. An unapodized spectral resolution of 1 cm$^{-1}$ was retained, with an iris aperture diameter of 3 mm. A 2.5 m long multi-pass cell, with 60 μm thick polypropylene windows, was connected to the FTS and used with an optical path of 151.75 m. All measurements were made at room temperature and 1000 scans were co-added to yield each spectrum in the 40-680 cm$^{-1}$ spectral range. The experiments were carried out as follows. Purified water vapor was first introduced inside the cell (at a pressure measured with a 0-10 mb gauge) and $CO_2$ was then added up to a total pressure of about 1 bar (measured with a 0-1 bar gauge). After a delay of about one hour (for the homogenization of the mixture) a first spectrum was recorded. The cell was then evacuated step by step and a series of spectra were collected for progressively reduced total pressures. As in [7], reference (100% transmission) spectra were obtained for similar pressures with pure argon gas inside the cell. In these experiments, total pressures between about 0.1 and 1 bar were investigated, for $H_2O+CO_2$ mixtures containing from 0.04 to 1 % of water vapor.

*2.2 Analysis*

Transmission spectra were first obtained by dividing each gas sample spectrum by a reference one (empty cell or argon gas). Let us emphasize that, since the targeted continuum absorptions are weak and vary slowly with wave number, precise knowledge of the 100% transmission level is crucial. With this purpose, the LISA evacuated cell spectrum at each temperature was corrected for the eventual mechanical influence of the presence of gas inside the cell. This was done by multiplying it by a constant such that it matches the pure $CO_2$ spectrum at the same temperature near 1150 cm$^{-1}$ where the $CO_2$ absorption is negligible. For the experiments made at SOLEIL, as done in [7], the argon spectrum recorded for the closest pressure was retained.

The total absorption coefficient (i.e. $\alpha$ in cm$^{-1}$) at wave number $\sigma$ (cm$^{-1}$) of a $CO_2+H_2O$ mixture at temperature $T$, of total density $\rho_{Tot}$ and mole fractions $x_{CO_2}$ and $x_{H_2O}$ can be written as:

$$\alpha(\sigma,T,\rho_{Tot},x_{CO_2},x_{H_2O}) = \sum_{A=CO_2,H_2O} \alpha^A_{Local}(\sigma,T,\rho_{Tot},x_{CO_2},x_{H_2O}) \\ + \sum_{A=CO_2,H_2O} \sum_{P=CO_2,H_2O} \alpha^{A-P}_{CA}(\sigma,T,\rho_{Tot},x_{CO_2},x_{H_2O}) \quad , \qquad (1)$$

where $\alpha^A_{Local}$ denotes the absorption due to local lines of the absorbing species 'A' whose extensions are limited to $\pm\Delta\sigma_A$ around the line center and $\alpha^{A-P}_{CA}$ is the continuum absorption due to absorber A interacting with the perturber 'P'. Provided that $\Delta\sigma_A$ is much greater than the pressure-broadened widths of the lines of species A under the considered sample conditions, one can write [2,3]:

$$\alpha^{A-P}_{CA}(\sigma,T,\rho_{Tot},x_{CO_2},x_{H_2O}) = x_A x_P \rho^2_{Tot} C_A^{A-P}(\sigma,T) \quad , \qquad (2)$$

where $CA^{A-P}(\sigma,T)$ is the squared-density-normalized continuum absorption due to molecule A "perturbed" by the presence of molecule P.

In order to deduce $CA^{H_2O-CO_2}(\sigma,T)$ from the measured spectra, the following three-steps were used.

1- $\alpha_{Local}^{H_2O}$ and $\alpha_{Local}^{CO_2}$ were first calculated by using spectroscopic data from the HITRAN 2012 database [8], complemented with H$_2$O-broadening coefficients of CO$_2$ lines calculated using the analytical expressions proposed in [9] and CO$_2$-broadening coefficients of H$_2$O lines from [10]. When not available in databases (e.g. the pressure shifts of CO$_2$ lines induced by collisions with H$_2$O at room temperature) the collisional parameters were assumed to be the same as in air or neglected (e.g. the temperature dependence of the shifts). The consequences of these necessary approximations are difficult to evaluate but are expected to be practically negligible considering the small values of the line shifts and spectral resolution of the present measurements. The influence of the FTS finite resolution was taken into account by convolving the associated calculated spectral transmissions (i.e. $\exp[-\alpha_{Local}^{H_2O}(\sigma)L]\exp[-\alpha_{Local}^{CO_2}(\sigma)L]$ with $L$ the path length) with an instrument line shape taking the maximum optical path difference and the iris radius into account. The absorption by each H$_2$O line was calculated between -25 and 25 cm$^{-1}$ from the line center (i.e. with $\Delta\sigma_{H_2O}$ = 25 cm$^{-1}$) using a Voigt profile, and the "pedestal" value at 25 cm$^{-1}$ (see [2] and Sec. 3) was subtracted. This was done for consistency with the choice adopted for the widely used MT_CKD self and foreign continua $CA^{H_2O-H_2O}(\sigma,T)$ and $CA^{H_2O-N_2}(\sigma,T)$ [2,11]. For CO$_2$ lines, $\Delta\sigma_{CO_2}$ = 5 cm$^{-1}$ was used in the computation of $\alpha_{Local}^{CO_2}$ using Voigt line shapes for consistency with the χ-factor corrections proposed in [1,12] (see below).

2- In a second step, the CO$_2$- and H$_2$O-broadened absorptions by the wings of CO$_2$ lines, $CA^{CO_2-CO_2}(\sigma,T)$ and $CA^{CO_2-H_2O}(\sigma,T)$, were calculated using the $\chi$ factors of [12] and [1], respectively, and also removed from the experimental transmissions. The use of these factors, which were obtained in the high frequency wing of the CO$_2$ $\nu_3$ band above 2400 cm$^{-1}$ to model the sub-Lorentzian behavior of the $\nu_2$ band lines can introduce uncertainties [since it is known that the $\chi$ factors are different in different spectral regions (e.g. [13,14])]. However, as shown in [1], the use of several different line wings leads to very small differences in computed transmissions for path length and pressure conditions close to those of the present work. The consequences of uncertainties on the used CO$_2$ wings model on the determination of the CO$_2$-broadened H$_2$O continuum should thus be very small.

3- Finally, the contribution of the pure H$_2$O continuum $CA^{H_2O-H_2O}(\sigma,T)$ was calculated using the MT_CKD 3.0 parameterization [2] (available at http://rtweb.aer.com/continuum_frame.html ) and removed from measured values. This exercise provides, from each measured spectrum, the values of $\alpha_{Wing}^{H_2O-CO_2}$, i.e. of $x_{H_2O}x_{CO_2}\rho_{Tot}^2 CA^{H_2O-CO_2}(\sigma,T)$ (that includes the 25 cm$^{-1}$ pedestal) at all wave numbers of the recorded spectra.

In order to test the consistency of our measurements and data analysis procedure, we looked at how well the retrieved values of $\alpha_{\text{Wing}}^{H_2O\text{-}CO_2}$ verify Eq. (2). As could be expected, very large uncertainties affect the results for the numerous spectral points where the relative contribution of $\alpha_{\text{Wing}}^{H_2O\text{-}CO_2}$ is very small. It is the case in regions close to the intense lines of $H_2O$ and $CO_2$ where $\alpha_{\text{Local}}^{H_2O}$ or $\alpha_{\text{Local}}^{CO_2}$ can be several orders of magnitude greater than $\alpha_{\text{Wing}}^{H_2O\text{-}CO_2}$. It is also the case in the near wing of the $CO_2$ $\nu_2$ band. We thus only retained, for the final results, the wave numbers where the relative contribution of the $CO_2$-broadened $H_2O$ wings is significant (i.e. in the troughs between $H_2O$ lines and sufficiently far away from the strong $CO_2$ absorption features). In order to carry this selection, we first removed the wave numbers in or close enough to line centers by ensuring that absorption is not saturated for any experimental pressure considered. The saturation condition was arbitrarily defined as absorption larger than 95% (or transmission lower than 5%). We then removed manually the wave numbers within ~3 cm$^{-1}$ of main, non-saturated lines. The remaining wave numbers are located in the troughs between $H_2O$ lines. In a second step, we averaged the absorption between $H_2O$ lines, with a typical wave number bin size of ~5 cm$^{-1}$ (thus much smaller than the 25 cm$^{-1}$ cut-off distance). With this procedure, the results obtained from the two experiments quite well verify Eq. (2), as illustrated by Fig. 1, and a posteriori validate our analysis.. The slope of the linear fits vs $x_{CO_2} x_{H_2O} \rho_{\text{Tot}}^2$ provide the final value of $CA^{H_2O\text{-}CO_2}(\sigma, T)$ which are given in Table 1 of the Supplementary Material file where the error bars correspond to twice the statistical uncertainty of the linear fits. Note that they do not take into account possible biases due to errors in the $H_2O$ self continuum [2] which was removed from the measured absorptions and makes a significant contribution in the experiments carried at LISA.

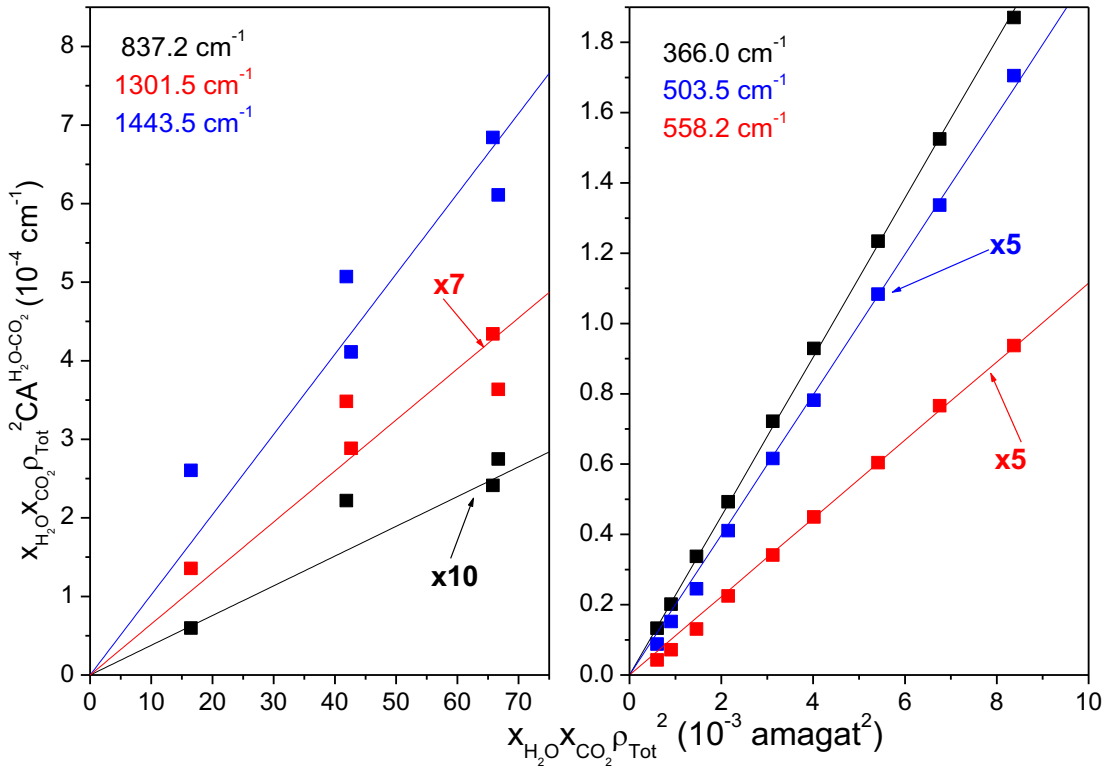

*Fig. 1: Experimentally determined $CO_2$-broadened $H_2O$ continuum absorptions (symbols) deduced from measurements carried at LISA (left panel) and SOLEIL (right panel) for selected wave numbers and their linear fits versus the product of the $H_2O$ and $CO_2$ densities.*

## 3. Calculations of the $CO_2$-broadened $H_2O$ continua

For comparisons with the experimentally determined values of the $CO_2$-broadened $H_2O$ continuum, calculations we carried using the only model available. The latter results from quasistatic predictions of the shape of the $CO_2$-broadened $H_2O$ line wings that were made nearly thirty years ago [5] and have not been tested so far. For consistency with the experimental results, the continuum absorption is written as:

$$CA_\chi(\sigma,T) = \alpha_\chi^{\text{Total}}(\sigma,T) - \left[\alpha_L^{\text{Local}}(\sigma,T) - \text{pedestal}(\sigma,T)\right], \quad (3)$$

where the total absorption coefficient is given by:

$$\alpha_\chi^{\text{Total}}(\sigma,T) = \frac{n_0}{\pi} \sum_{\text{all lines }\ell} S_\ell(T) \frac{\sigma \sinh(\text{hc}\sigma/2k_BT)}{\sigma_\ell \sinh(\text{hc}\sigma_\ell/2k_BT)}$$
$$\times \left[\frac{\Gamma_\ell(T)}{(\sigma-\sigma_\ell)^2 + \Gamma_\ell(T)^2}\chi(\sigma-\sigma_\ell,T) + \frac{\Gamma_\ell(T)}{(\sigma+\sigma_\ell)^2 + \Gamma_\ell(T)^2}\chi(\sigma+\sigma_\ell,T)\right]. \quad (4)$$

with $\chi(\sigma+\sigma_\ell,T) = \chi^-(\sigma+\sigma_\ell,T)$, $\chi(\sigma-\sigma_\ell,T) = \chi^+(\sigma-\sigma_\ell,T)$ for $\sigma-\sigma_\ell \geq 0$ and $\chi(\sigma-\sigma_\ell,T) = \chi^-(\sigma-\sigma_\ell,T)$ for $\sigma-\sigma_\ell \leq 0$. These expressions come from Eq. (40) of Ref. [5] in which the functions $\hat{\chi}^\pm(\sigma,T)$ (see Fig. 4 of [5]), which are in units of cm$^{-1}$/atm, were replaced by $\chi^\pm(\sigma,T)\Gamma_\ell(T)$ were $\chi^\pm(\sigma,T)\Gamma_\ell(T)$ is dimensionless and given by $\chi^\pm(\Delta\sigma,T) = \hat{\chi}^\pm(\Delta\sigma,T)/\hat{\chi}^\pm(\Delta\sigma = 0,T)$. This ensures that the absorption near a line center is a sum of the two Lorentzian profiles associated with the positive and negative resonances (Van Vleck-Weisskopf profile). The local and pedestal contributions in Eq. (3), also used in the treatment of measured spectra in order to deduce the continuum (see Sec. 2.2), are given by:

$$\alpha_L^{\text{Local}}(\sigma,T) = \frac{n_0}{\pi} \sum_{\text{lines }\ell \text{ with }|\sigma-\sigma_\ell|<25 \text{ cm}^{-1}} \frac{S_\ell(T)\Gamma_\ell(T)}{(\sigma-\sigma_\ell)^2 + \Gamma_\ell(T)^2}, \quad (5)$$

$$\text{pedestal}(\sigma,T) = \frac{n_0}{\pi} \sum_{\text{lines }\ell \text{ with }|\sigma-\sigma_\ell|<25 \text{ cm}^{-1}} \frac{S_\ell(T)\Gamma_\ell(T)}{(25)^2}. \quad (6)$$

In Eqs. (3)-(6), $n_0 = 2.687 \times 10^{19}$ molec/cm$^3$ is the number density of molecules at 1 amagat, $S_\ell(T)$ is the integrated intensity of the $H_2O$ line $\ell$ [in units of cm$^{-2}$/(molec.cm$^{-3}$)] and $\Gamma_\ell(T)$ is its $CO_2$-broadened HWHM for a density of 1 amagat [i.e. a pressure of $P(\text{atm}) = 273.15/T(\text{K})$].

In order to compute the values of the absorptions from these equations, the needed data were taken from [8] for the $H_2O$ line positions and integrated intensities, from [10] for the pressure broadening coefficients of $H_2O$ lines by $CO_2$, and from Fig. 4 of [5] for $\hat{\chi}^\pm(\sigma, 296\text{ K})$. Note that, in the absence of relevant information, we assumed $\hat{\chi}^\pm(\sigma,T)$ temperature independent in the relatively narrow temperature range of the experiments (from 295 K to 366 K).

## 4. Results and discussion

All available experimental values of the $H_2O$-$CO_2$ continuum are displayed in Fig. 2 together with predictions made at 296 K despite the fact that the measurements of [6] and those made at LISA have been carried at slightly higher temperatures. This is justified since both experiments showed no detectable dependence of the continuum on temperature. The results in Fig. 2 call for several remarks: (i) The first is that the various experimental determinations of the continuum are consistent if uncertainties are taken into account as shown by the present measurements and those of [6] around 1300 cm$^{-1}$. The second is that calculations using the line shape correction factors χ of [5] lead, on overall, to very good results if one considers that it was predicted nearly three decades ago without use of any experimental information and that the continuum varies by three orders of magnitude within the considered spectral region. This *a posteriori* validates these predicted $H_2O$ line-shape factors and the subsequent implications for Venus atmosphere discussed in [15] and for magma ocean atmospheres studied for example in [16]. Figure 2 however shows that calculations may slightly underestimate the absorption in the far wing (e.g. around 1200 cm$^{-1}$) and increasingly overestimate it when getting closer to the intense lines of the rotational and $\nu_2$ bands of $H_2O$ (below 400 cm$^{-1}$ and above 1400 cm$^{-1}$). Now, recall (Fig. 4 of [5]) that $\chi^{\pm}(\Delta\sigma, 296\,K) = \hat{\chi}^{\pm}(\Delta\sigma, 296\,K) / \hat{\chi}^{\pm}(\Delta\sigma = 0, 296\,K)$ is unity for $\Delta\sigma=0$ cm$^{-1}$ and then increases with $\Delta\sigma$ to reach a maximum of about 4 at around $\Delta\sigma=80$ cm$^{-1}$. It then decreases down, reaches back unity near $\Delta\sigma=250$ cm$^{-1}$ before a quick fall-off bringing it down to $3\times10^{-3}$ around 1100 cm$^{-1}$. The results in Fig. 2 indicate that the predicted line shape is too much super-Lorentzian at intermediate distance [i.e. $\chi^{\pm}(\Delta\sigma, 296\,K)$ goes too far above unity below 250 cm$^{-1}$] while it is not enough sub-Lorentzian in the farther wing (i.e. $\chi^{\pm}(\Delta\sigma, T)$ decreases too slowly above 250 cm$^{-1}$).

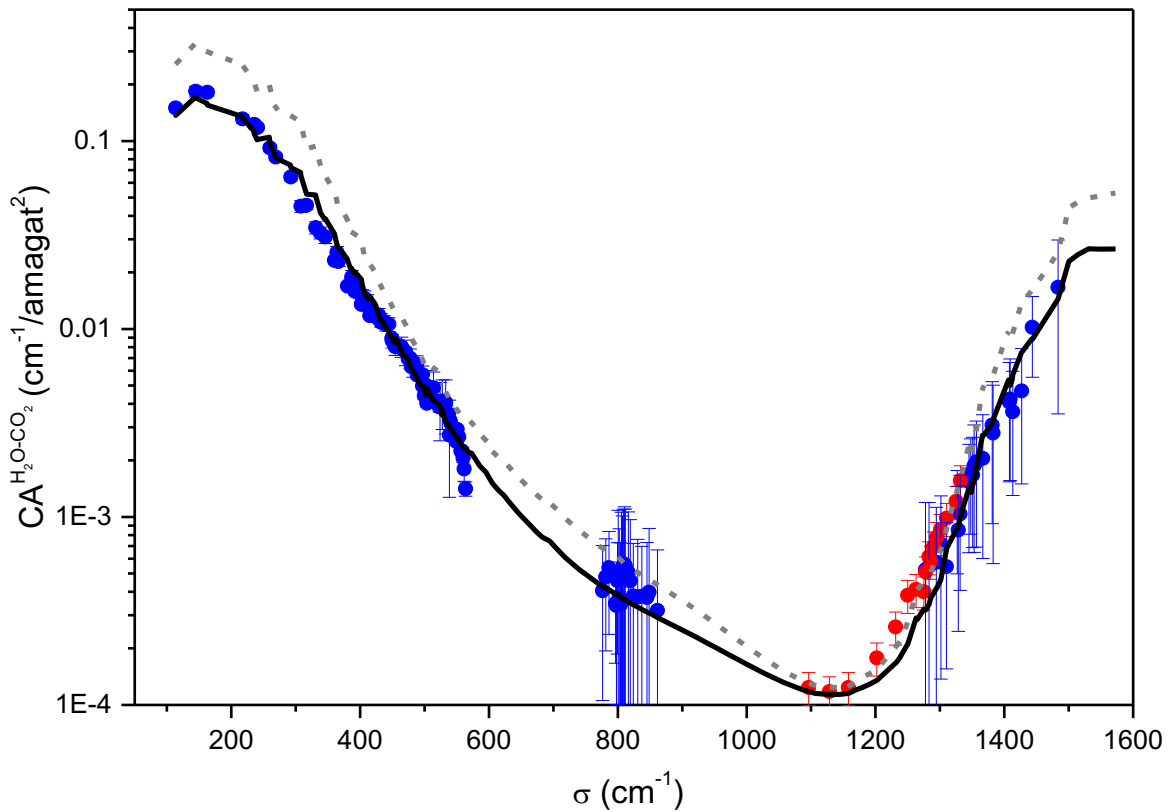

*Fig. 2: Water-carbon dioxide density normalized continua. The symbols denote values deduced from the experiments carried by the present study (blue) and those digitized from fig. 4 of [6] (red with indicative 20% error bars). The lines are results of calculations (see text)*

*carried using with the original χ factors of [5] (dashed line) and the empirically adjusted one (solid line).*

Accurate description of the $CO_2$-broadened wings of $H_2O$ lines is required to properly model $CO_2$-dominated planetary atmospheres such as Venus, ancient Mars, and exoplanets that are on the verge to be characterized. It is thus relevant to try to propose for this the best line shape. Empirical corrections to the $\hat{\chi}^{\pm}$ factors proposed in Ref. [5] have thus been made in order to improve the agreement with measured values. The obtained $\chi^{\pm}$ factors are given in Table 2 of the Supplementary Material file and plotted in Fig. 3 where they can be compared with the original ones. As can be seen in Fig.2 the agreement with the measured values of the continuum is significantly improved when the optimized line shape corrections are used.

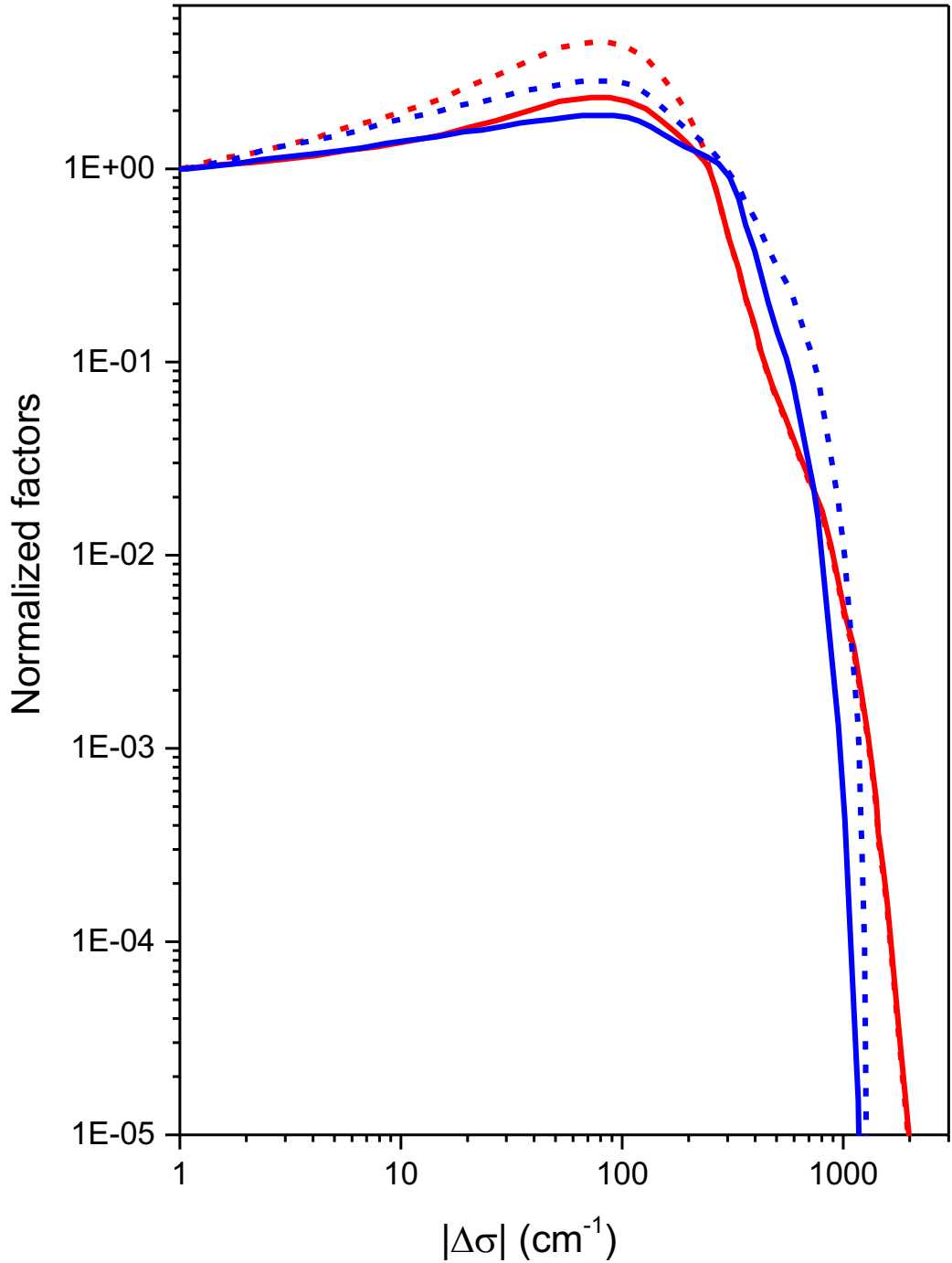

*Fig. 3: Room temperature CO$_2$-broadened H$_2$O line shape correction functions $\chi^+$ (red) and $\chi$ (blue) normalized to unity at =0 cm$^{-1}$ from Ref. [5] (dashed lines) and corrected for better agreement with the measured continuum (solid lines).*

### 5. Conclusion

Values of the CO$_2$-broadened continuum absorption by H$_2$O between 100 and 1500 cm$^{-1}$ have been determined from transmission measurements made using two different experimental setups. They are in good agreement with previous experimental results and *a posteriori* demonstrate that direct theoretical predictions made a long time ago using a quasistatic approach lead to quite satisfactory results. In order to correct for the remaining discrepancies between the measured and calculated continua, *ad-hoc* line shape corrective factors have been adjusted on the available experimental values. The resulting model should be useful for calculations of the contribution of water vapour to radiative transfer in $CO_2$-rich atmospheres. However, complementary studies are still needed in order to test the model at higher wavenumbers (eg in the 5 m window on the high frequency side of the $H_2O$ $_2$ band) which are also important for planetary radiative transfer and, if possible, other temperatures. For this, new experimental investigations must be made, which we plan to carry in the future.

**Acknowledgements**
The authors are grateful to R. Gamache for providing the data for the broadening of $H_2O$ lines by $CO_2$. They also thank O. Pirali and the SOLEIL-AILES beam-line staff for their technical support during the far infrared measurements and L. Manceron for lending a pressure transducer.